\documentstyle{article}   
\newcommand{\bibi}{\bibitem}                                                  
\newcommand{\etal}{\it {et al.}}

\newcommand{\half}{\frac {1}{2}}                                             
                                             
\newcommand{\beq}{\begin{equation}}                                           
\newcommand{\eeq}{\end{equation}\noindent}                                  
\newcommand{\beqr}{\begin{eqnarray}}                                          
\newcommand{\eeqr}{\end{eqnarray}\noindent}                                   
\newcommand{\vQ}{\bf Q}                                                      
\newcommand{\vr}{{\bf r}}                                                    
\newcommand{\vk}{{\bf k}}                                                     
                                                     
\newcommand{\vq}{{\bf q}}

\begin{document}     
\title{
On Possible Coexistence of Superconductivity and Charge Density Wave 
in Hole-doped C$_{60}$} 

\author{Sanjoy K. Sarker \\                                                   
Department of Physics,                                        
The University of Alabama, Tuscaloosa, AL 35487 \\}                           
\maketitle 
\begin{abstract} 

Superconductivity in C$_{60}$ materials is modeled in terms of
an intramolecular, nonretarded attraction.        
It was shown previously that, at intermediate coupling, 
the model possesees a state in which s-wave
superconductivity coexists with a charge density wave, the latter
stabilized by intersite repulsion. The CDW
causes $T_c$ to decrease near half filling. We argue that
hole-doped C$_{60}$, for which $T_c$ peaks away from half filling,
is a possible candidate for this state.      
But the electron-doped C$_{60}$ and A$_3$C$_{60}$ are conventional
superconductors, stabilized against a CDW by metallic screening,
which is treated in a parameter-free fashion.

                               
\end{abstract}

About a decade ago, superconductivity with surprisingly high
$T_c$'s ($30-40K$) was observed in A$_3$C$_{60}$, where A is an 
alkali atom \cite{heb,ram}. Recently much higher $T_c$s ($52K - 117K$)
have been seen in hole-doped C$_{60}$ \cite{sch,sch2}.
Within the Bardeen-Cooper-Schrieffer (BCS) picture,
origin of such high $T_c$'s has been attributed to special properties      
C$_{60}$, such as, high density of states (DOS) and large phonon 
frequency. However, variation of $T_c$ with the density $n$
does not follow the DOS. Also,
in the  electron-doped C$_{60}$ and A$_3$C$_{60}$ the 
maximum of $T_c$ occurs near half filling. But in the hole-doped 
material, $T_c$ peaks at a density of $n = 3-3.5$ holes/molecule, 
and decreases rapidly toward half filling (5 holes/molecule) 
\cite{sch}. Such a behavior has been predicted previously
on theoretical grounds\cite{sar1,sar2}. The sharp 
decline in $T_c$ is caused by the appearance of a charge density 
wave (CDW) near half filling which, due to the frustration effects 
of the lattice, can coexist with superconductivity. Here we
examine the conditions for the existence of such a state in 
C$_{60}$. 

The possibility of a CDW instability in C$_{60}$ was 
recognized early by Zhang, Ogata and Rice\cite{zor}. As discussed
below, it becomes more likely as $T_c$ increses. The key issue is 
the effect of Coulomb repulsion which opposes superconductivity, 
but favors the CDW. We will focus mostly on the normal state, and
consider realistic potentials. We find that if there is 
no metallic screening, the energy gained by the CDW increases
monotonically with increasing interaction strength. However,
in the presence of metallic screening, the gain has 
a {\em maximum}. It is too small to stabilize the CDW state in the lower 
$T_c$ materials such as A$_3$C$_{60}$. But for the hole-doped
case, the CDW state can not be ruled out. We discuss
some normal-state anomalies of this state.     

Crystalline $C_{60}$ has a face-centered-cubic (FCC) structure,
and is a semiconductor. The conduction band is
formed from the 3-fold degenerate $t_{1u}$ molecular
orbitals, and the valence band from the 5-fold degenerate $h_{1u}$ 
orbitals. In A$_3$C$_{60}$, each alkali atom donates one 
electron to the conduction band making it half-filled.
Large molecular size leads to a large lattice spacing 
($a \sim 10 \AA$), and consequently very narrow bands \cite{erw} 
of width $W \sim 0.5 eV$, which can thus be adequately described 
by a tight-binding model with nearest-neighbor hopping. This 
implies that geometric effects such as {\em frustration}
and {\em nesting} are important. 

Our theory is based on two key assumptions. (1) Low energy physics of 
the doped system can be described by an effective Hamiltonian defined 
entirely within the relevant band, i.e., the $h_{1u}$ band for
hole-doped C$_{60}$, and the $t_{1u}$ band for electron-doped C$_{60}$
and A$_3$C$_{60}$.  The effect of integrating out other bands and 
phonon degrees of freedom is to renormalize the interactions.
(2) The attractive interaction responsible for 
superconductivity is intramolecular in origin and 
effectively {\em nonretarded}, i.e, the characteristic 
frequency $\omega _0$ is comparable to the Fermi energy. Indeed,
intramolecular phonons that are thought to be responsible \cite{sch}
have rather high frequencies: $\hbar\omega _0 \sim 0.15 eV - 0.2 eV$, 
easily comparable with the Fermi energy. 

The model always has a solution in which the normal state is
conventional, and superconductivity is BCS like.  
In the BCS theory $T_c$ is given by
\beq kT_c \approx 1.3 \hbar\omega _0 e^{-1/\lambda}, \eeq
where $\lambda \approx U\rho (\epsilon _F)$ is the dimensionless 
coupling constant, $\rho (\epsilon _F)$ is the density of states (DOS) 
at the Fermi level, $U$ is the attractive interaction. 
The observed high $T_c$'s are usually\cite{sch} attributed to:
(1) a relatively high DOS due to the narrow bandwidth ($\rho \propto
W^{-1}$) and orbital degeneracy, and (2) a large prefactor due to 
high phonon frequency. Furthermore, in A$_3$C$_{60}$, $T_c$ 
increases with increasing size of the alkali atoms. This is
because as the lattice expands, $W$ decreases.
Similarly, the five-fold degeneracy of the $h_{1u}$ level 
should lead to higher DOS \cite{erw}, 
which will account for the higher $T_c$ in the hole-doped material.

However, not everything fits. According to the simple BCS formula,
the $n$-dependence of $T_c$ should track that of the DOS
(through the $n$-dependence of the Fermi energy). This is actually 
not the case. For the FCC lattice
the DOS has sharp features. The measured $T_c$ is
smoother, its maximum is far removed from that of the DOS, both
for the electron doped or hole doped materials\cite{sch}.

We point out that the problem is partly due to the assumption 
that the attractive interaction is retarded, i.e., 
confined within $\hbar\omega _0 << \epsilon _F$ of the Fermi surface. 
Then one can approximate the DOS by $\rho (\epsilon _F)$, and use
$\hbar\omega _0$ as a cut-off in the energy integral which accounts
for the prefactor. This procedure is not valid when $\hbar\omega _0$ 
is comparable to $\epsilon _F$. In fact, as shown previously\cite{sar2}, 
in the nonretarded case $T_c$ varies much more smoothly with $n$. The 
reason is that for general values of $\hbar \omega _0$,  
a more accurate  approximation is given by  
$$ kT_c \approx 1.3 \hbar(w_1w_2)^{1/2}
e^{-1/\rho _0U}, $$
where $\hbar w_1 = min(\epsilon _F, \hbar\omega _0)$, 
$\hbar w_2 = min(W - \epsilon _F,\hbar\omega _0)$, and $\epsilon _F$
is measured from the bottom of the band. Here $\rho _0$
is the weighted average of the DOS $\rho (\epsilon)$  
over the allowed region with a slowly varying weight factor 
$\epsilon ^{-1}\tanh (\epsilon/2kT_c)$ (normalized to unity).
For small $\hbar \omega _0$ we recover the usual BCS formula,
whereas for large $\hbar\omega _0$, 
the variation of $T_c$ with $n$ is smoothed out considerably.
In this case, a better estimate is $\rho _0 \approx 1/W$.   

A more important issue is the possibility of a CDW instability
which can not be ignored when $\hbar\omega _0$ is 
comparable to $\epsilon _F$. We consider a one-band model,
characterized by a single-particle energy $\epsilon _{\vk}$, 
an on-site attraction $U$, and an intersite repulsive interaction 
$V(\vr)$. The relevant 
dimensionless energy scales are $U\rho$ and $V\rho$. 
The effect of orbital degeneracy 
can be approximately taken into account by changing $\rho$, or, 
equivalently, by changing $U/W$ and $V/W$. The possibility of a
CDW was studied in reference \cite{zor} using this model, but
without the $V$ term. However, as shown previously,
inclusion of even a nearest neighbor $V$ leads to nontrivial
consequences \cite{sar1,sar2}.  We now generalize to 
$V(\vr)$ of arbitrary range.

In the simplest mean-field approximation, and in the absence of
superconductivity, a CDW with a wavevector 
$\vQ$ would appear for \beq (U + 2V_{\vQ})\chi (\vQ) > 1, \eeq 
where $\chi (\vQ)$ is the charge susceptibility for the
noninteracting system, and 
$V_{\vQ} = - \sum _{\vr \neq 0}e^{i\vQ.\vr}V(\vr)$.  
The minus sign is included in order to make $V_{\vQ}$ positive. 
For a bipartite (e.g., simple cubic, BCC etc) lattice, 
there is a $\vQ$ for which
\beq \epsilon _{\vk + \vQ} = - \epsilon _{\vk}\eeq 
for all $\vk$. Then the Fermi surface is perfectly nested at
half filling, leading to a logarithmically divergent $\chi $
at $T =0$ and a CDW  instability for any $(U + 2V_{\vQ}) > 0$. 
For the nonbipartite FCC lattice, $\chi (\vQ)$ is finite
{\em even at half filling} as there is no nesting, i.e., 
no $\vQ$ for which Eq.(3) holds. A conventional metallic state is 
then stable for $(U + 2V_{\vQ}) < 1/\chi (\vQ)$ 
$\approx 0.375W$. Above this, a CDW state appears with 
${\vQ} = (0,0,2\pi/a)$, which corresponds to a CDW with a density 
$n(\vr) = n + n_1e^{i\vQ.\vr}$, i.e., 
on successive planes perpendicular to the $z$ axis, the density
alternates between $n + n_1$ and $n - n_1$. 

From (2), it appears that a CDW can exist even $V = 0$. 
However, such a state is unstable against s-wave superconductivity,
since the pair susceptibility is logarithmically divergent. 
The inclusion of $V(\vr)$ changes this picture, due to two reasons. 
First, the $V$ term favors the CDW energetically.
At the mean-field level, the excess potential energy per site
in the CDW state equals $-(U + 2V_Q)n_1^2/4$. 
Physically,  the CDW forms an {\lq\lq ionic crystal"} with
$V_{\vQ}$ proportional to the
appropriately generalized Madelung energy\cite{ash1}.
However, compared with a bipartite (e.g., BCC) lattice, 
charge ordering on 
an FCC lattice is {\em frustrated}, leading to
a somewhat lower energy gain. Nonetheless, 
we have verified numerically that $V_Q > 0$,
for the screened Coulomb potentials (Fig 1) which are the
potentials of interest.  Second, $V(\vr)$ always opposes 
superconductivity by keeping particles apart,  
stabilizing the CDW state at larger $V_{\vQ}$.

The CDW splits the noninteracting band into two bands of energy   
\beq E_{\pm}(\vk) = \epsilon _1(\vk) \pm 
\lbrack \epsilon _2^2(\vk) + \Delta _C^2\rbrack ^{\half}, \eeq
where 
$\epsilon _1(\vk) = \half (\epsilon _{\vk} + \epsilon _{\vk+\vQ})$
and  
$\epsilon _2(\vk) = \half (\epsilon _{\vk} - \epsilon _{\vk+\vQ})$.
At the same $\vk$, the bands are separated by a gap
$\Delta _C = n_1(U + 2V_{\vQ})/2$. For a  bipartite lattice the 
geometrical property, Eq. (3), ensures that $\epsilon _1(\vk) = 0$, 
leading to insulating behavior. However, Eq.(3) does not hold for the 
FCC lattice. Then for small enough $\Delta _C$,  
the maximum of the lower band lies above the minimum of the upper
band which, in general, is at a different $\vk$ point. This leads to
a semimetal with a partially gapped Fermi surface. 
The is a direct consequence of
geometric properties frustration and (lack of) nesting. Eventually, there
is a transition to a an insulating state for $(U + 2V_{\vQ}) > 0.562W$.

Furthermore, residual attractive interaction within the semimetal 
leads to superconductivity. For a nearest-neighbor $V$, the CDW 
has been shown to coexist with superconductivity over a substantial 
region in the parameter space \cite{sar1}.
 The presence of the CDW lowers $T_c$, and yields \cite{sar2}
a $T_c(n)$ which is very different from that
in the conventional state (no CDW). In either case, $T_c$ does not
track the DOS. In the conventional state, it peaks
near half filling and gradually decreases to zero at the band edges.
This is similar to the behavior observed in electron doped C$_{60}$
and A$_3$C$_{60}$.
 
By contrast, in the CDW-superconductor $T_c$ has a {\em minimum} near 
half filling where the CDW instability is strongest. Away from half
filling, $T_c$ increases as the CDW weakens. Eventually,
there is a transition to the conventional state.
Consequently $T_c$ reaches a maximum near the 
transition point and then decreases following the behavior in the
conventional state. This is very similar to the behavior observed
in the hole doped $C_{60}$. So far only one maximum has been seen
because experiments have been limited to only one side of 
half filling. The theory predicts two, one on each side.
These results should be qualitatively correct for the
more realistic screened Coulomb potential, although one expects 
a more rapid decline of $T_c$ toward the band edges since screening
length increases with distance between particles.                       

Are the observed $T_c$'s consistent with the theory? Consider 
the conventional state at half filling which
exists for $(U + 2V_{\vQ})/W < 0.375$. To estimate $U/W$ from 
the measured $T_c$, we use formula (1), conveniently rewritten as
$1/\lambda = - ln(kT_c/1.3\hbar\omega _0)$, to compute $\lambda$,
which is treated as a phenomenological coupling constant. At half filling, 
we can take $\omega _0$ to be the (intramolecular) phonon frequency,
which is same for all materials.  Noting that $\lambda$ varies slowly with
$\omega _0$, we set $\hbar\omega _0 = 1500K$, which is
reasonable. For the electron 
doped C$_{60}$ ($T_c = 11K$) this gives $\lambda = 0.146$.
However, for Rb$_3$C$_{60}$ ($T_c = 33K$), $\lambda = 0.245$.
Next step is to relate $\lambda$ with the theoretical parameter 
$U/W$. As discussed before, in the absence of $V(\vr)$, and
for large $\hbar\omega _0$, $\lambda \approx U/W$. However,
when $V(\vr)$ is included, $\lambda$ will be reduced so that 
$U/W > \lambda$. This means that electron-doped
C$_{60}$ and A$_3$C$_{60}$ can be conventional metals, but
for the larger alkali (Rb etc) materials $V_{\vQ}/W$ has to be
rather small.   

For the hole-doped C$_{60}$, the maximum $T_c$ of $52K$ yields 
a larger value of $\lambda$ ($\approx 0.28$) which could
be sufficient for the CDW state.  However, the
maximum is away from half filling which in our theory is due 
to the transition to the CDW-superconducting state. In order
to be consistent, we have to use the $T_c$
of the corresponding conventional state at half filling.
This is about 30 to 50 percent higher, 
which gives $U/W > \lambda \sim 0.32 - 0.34$. For any
reasonable values of $V_{\vQ}$ these materials should
be CDW metals near half filling. 
Any expansion of lattice will increase $U/W$, making
the CDW even more likely \cite{sch2}. 

The intriguing question is, why is A$_3$C$_{60}$ not a
CDW metal, as suggested previously\cite{sar1}? For example, 
for Rb$_3$C$_{60}$ to be conventional, $V_{\vQ}/W$ has to be 
$\sim 0.05$ or less, almost
two orders of magnitude smaller than the
bare Coulomb energy scale $e^2/a \approx 2.6 W$. 
Now, there are two sources of screening.
(1) Screening by other molecular bands 
which can be approximated by a static dielectric 
constant $\epsilon$. One expects $\epsilon$ to be somewhat larger
than that in ordinary semiconductors because of small (molecular) 
band gaps. (2) Since the transition is from one metallic state
to another, we must include {\em metallic} screening which
leads to exponentially short-range potential and much smaller 
$V_{\vQ}$. To see this, we consider the screened Coulomb potential 
\beq V(\vr) = \frac{e^2}{\epsilon r}e^{-r/\xi},\eeq 
where $\xi$ is the Thomas-Fermi screening length. 

Let us define the dimensionless quantity $\alpha$ by writing
$V_{\vQ} = v_0\alpha(\xi/a)$, where $v_0 = e^2/{\epsilon a}$
sets the energy scale for the long-range part.
Then $\alpha(x)$ is generalized Madelung constant \cite{ash1},
defined for screened-Coulomb potential. As shown in Fig.
(1) $\alpha$, which measures the energy gained by the CDW,
increases with the range of interaction $\xi/a$. In the absence of
metallic screening, $\xi = \infty$, and $\alpha \approx 1.6$.
With decreasing $\xi$, $\alpha$ decreases rapidly below
$\xi \sim a$, which is essentially the metallic region. Fig. 1
also shows that $\alpha$ is always smaller for the FCC lattice
than the bipartite BCC lattice. This is due to 
frustration which is another --- albeit smaller --
source of reduction of $V_{\vQ}$.

It is possible to have a CDW state in every case by choosing
the parameters $v_0$ (i.e., $\epsilon$) and $\xi$ appropriately.
However, a much better {\em parameter-free} understanding of 
screening can be obtained as follows. First, note that
$\xi$ itself depends on $v_0$. 
The Thomas-Fermi expression for $\xi$ \cite{ash2}, 
written in the lattice language, reads:
\beq \xi/a = \lbrack (4\pi\gamma v_0)\frac{dn}{d\mu}\rbrack ^{-1/2},\eeq 
where $\mu$ is the chemical potential, and the geometrical
parameter $\gamma = \sqrt 2$ for the FCC lattice. At low $T$, 
$dn/d\mu \approx \rho (\epsilon _F) \sim 2/W$. Let us define an
effective bandwidth by $2/W_{eff} = dn/d\mu$.  Since 
$\alpha$ depends only on $\xi/a$, we see that the dimensionless
quantity $V_{\vQ}/W_{eff}$ is a universal function of the
dimensionless coupling constant $v_0/W_{eff}$ only. 
This is plotted in Fig. 2.  Notice that
$V_{\vQ}/W_{eff} \sim V_{\vQ}/W$, the quantity of interest.
Naively one would expect $V_{\vQ}$ 
to keep increasing as $v_0$ increases. On the contrary,
$V_{\vQ}/W_{eff}$ has a broad maximum. This is because
an increased $v_0$ (i.e., a larger charge) makes the
screening length $\xi$ shorter (Eq. 6) which decreases $\alpha$.
The maximum occurs at $\rho v_0 \approx 0.15$ and has 
a value of about 0.042.  This is a pure number, which
depends only on the lattice type, but not any parameters.
Hence, we have the remarkable result
that irrespective of how large or small $v_0$ (i.e., $\epsilon$)
is, $2V_{\vQ}/W_{eff} < 0.084$. The smallness of this number,
combined with the above estimates of $U/W$, explains why
A$_3$C$_{60}$ is conventional. We stress that this
drastic reduction of $V_{\vQ}$ would not occur
if the CDW state were an insulator since
there is no metallic screening at the transition.

Since $V_{\vQ}/W_{eff}$ varies slowly (fig. 2),
its maximum value also
provides an estimate for its magnitude. For example,
$2V_Q/W_{eff} > 0.06$ in the entire range of 
$0.05 < v_0/W_{eff} <0.4$ ($52 > \epsilon > 6.5$). This
is sufficient for the hole-doped C$_{60}$ to be a CDW 
semimetal. This analysis is of course approximate. One
way to see if the state is experimentally realized
is to look for normal state anomalies. 

The semimetal is described in terms of two bands
separated by a direct gap $\Delta_C (T)$, as well as a
Fermi surface. The gap vanishes at $T_{CDW}$, above which the 
normal state is conventional.   
Below $T_{CDW}$, the band parameters become
temperature dependent through $\Delta _C(T)$,
leading to distinct features. Thus, the carrier density
becreases with increasing $\Delta _C$, leading to
peak or a shoulder in the resistivity as a function
of $T$. Other quantities, such as the specific heat,
also show similar behavior. To see this, consider the
approximate low $T$ expression for energy
\beq E(T) = E_0(\Delta _C)  
   + \frac {\pi ^2}{6}(kT)^2 \rho (\epsilon _F),
\eeq
where $E_{0}(\Delta _C)$ is the ground-state energy and
$\epsilon _F$ is the corresponding Fermi energy for a given 
$\Delta _C(T)$. For $T > T_{CDW}$, $\Delta _C = 0$, 
and $E_0$ and $\rho$ are independent of $T$,
leading to linear heat capacity and 
$T$-independent magnetic susceptibility. 
Below $T_{CDW}$, $E_0$ and $\rho (\epsilon _F)$ become 
temperature dependent through $\Delta _C(T)$. 
We have solved the mean-field equations numerically at 
finite $T$. Fig. 3 shows typical behavior of specific heat
as a function of $T/T_{CDW}$, which
is linear in $T$ above $T > T_{CDW}$, and again 
becomes linear at low $T$, as $\Delta _C(T)$ saturates,
but with a smaller slope, corresponding to
reduced value of $\rho (\epsilon _F)$. The sharp feature
below $T_{CDW}$ is due to the rapid decline in 
$\rho (\epsilon _F)$ as function of temperature, also
shown in Fig. 3. These features
should be experimentally observable in the region where
$T_{CDW} > T_c$, i.e., close to half filling.

In conclusion, we have shown that despite relatively high
$T_c$'s, frustration and metallic screening suppresses a
CDW instability in A$_3$C$_{60}$. But for hole-doped 
C$_{60}$, a mixed state can not be ruled out. If the anomalies
discussed above are not observed, then the CDW is absent.
This is possible, given the simplicity of the model and
approximate nature of the analysis. Even then the instability
can not be be very far away, and can occur
as $T_c$ is increased further by lattice expansion. The author
thanks C. Jayaprakash and T. L. Ho for discussions.

\bigskip

\centerline{\bf FIGURE CAPTIONS}

\noindent {\bf Fig.1:}
The generalized Madelung constant $\alpha = V_{\vQ}/v_0 =
\sum _{\vr \neq 0} \frac{e^{i\vQ.\vr
- r/\xi}}{(r/a)}$ plotted against dimensionless screening length $\xi/a$,
where $a$ is the nearest neighbor distance. Note that $\alpha$ depends
only on $\xi/a$ and the lattice type. The solid line is for the FCC lattice 
and the dashed line for the bipartite BCC lattice, showing that energy gain 
from the
CDW is less for the FCC lattice due to frustration. In usual metals,
$\xi/a < 1$.

\medskip
 
\noindent {\bf Fig.2:}
The dimensionless quantity $V_{\vQ}/W_{eff}$ is plotted 
(solid line) against
$v_0/W_{eff}$ where $W_{eff} = 2/\frac{dn}{d\mu} \sim W$. 
Also shown (dashed line) is $10\xi/a$ which, like $\alpha$, 
is slowly varying except for very small $v_0$. 

\medskip

\noindent {\bf Fig.3:} Scaled specific heat (solid line)
and density of states at the Fermi level (dashed line) are
plotted vs $T/T_{CDW}$ for the CDW-metal at half filling. 
The parameter values are $U + 2V_{\vq} = 6.71t$, $T_c=0.768t$, 
The prominent feature below $T_{CDW}$ is due to the CDW and
reflects the rapid rise in the DOS.


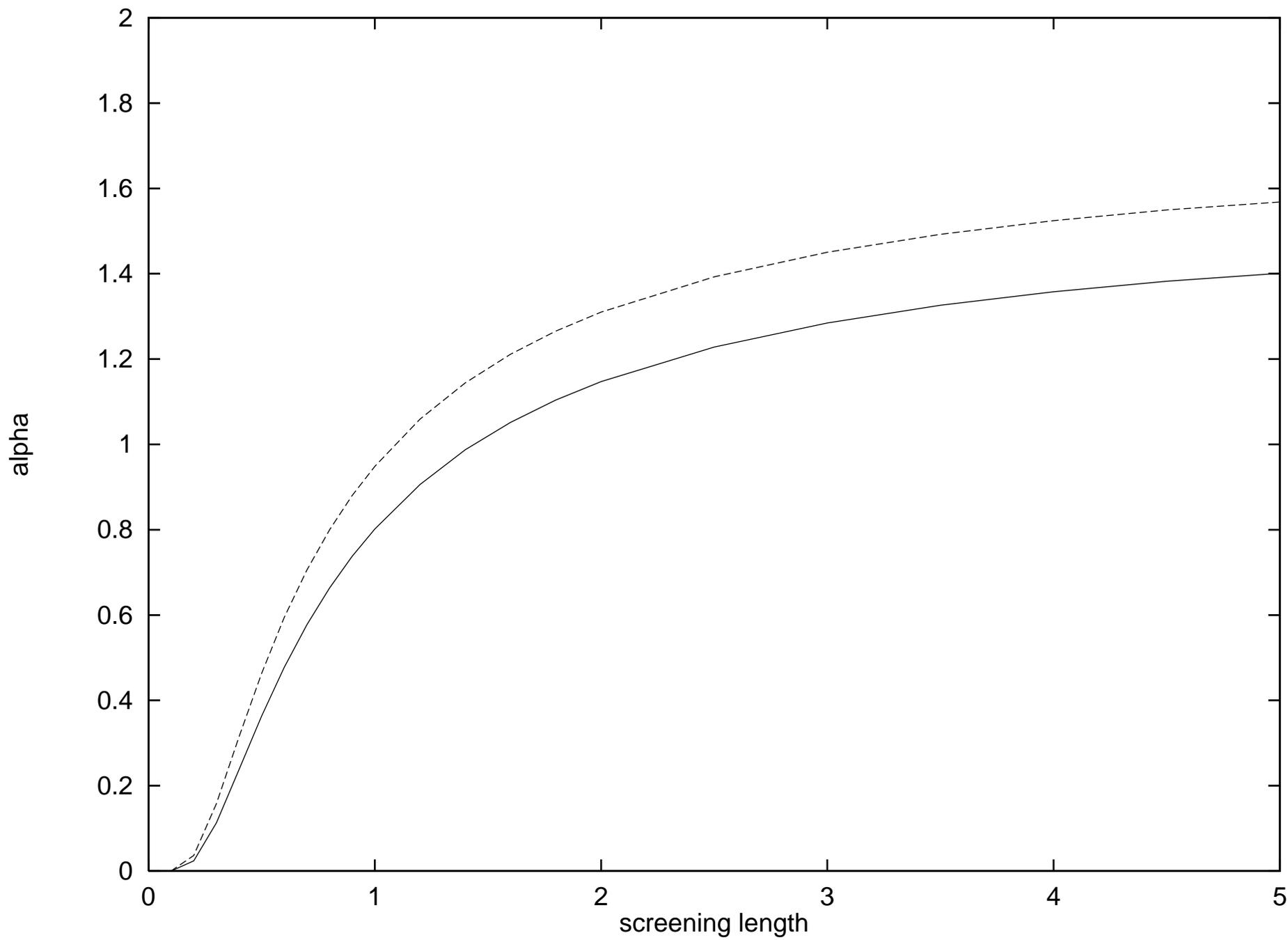
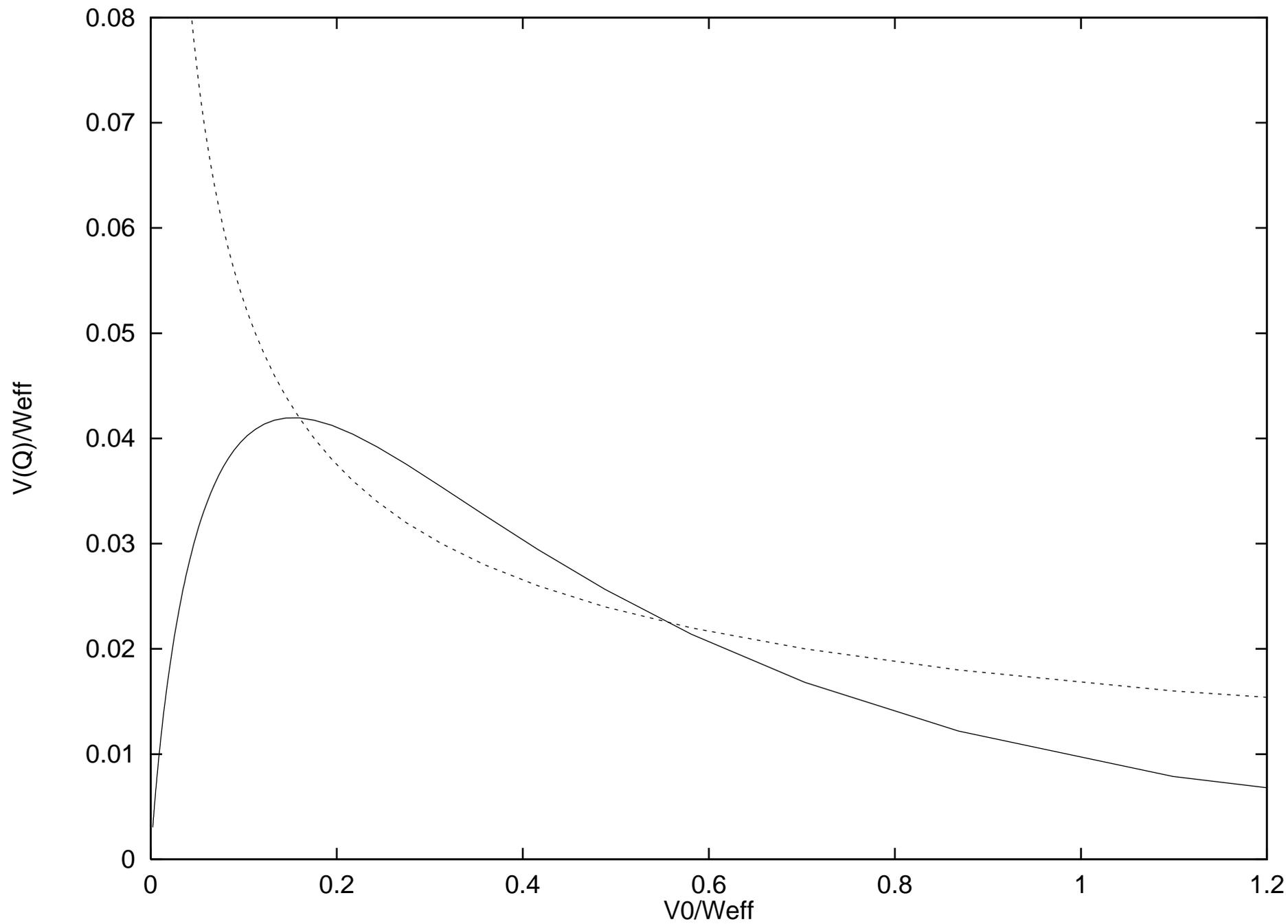
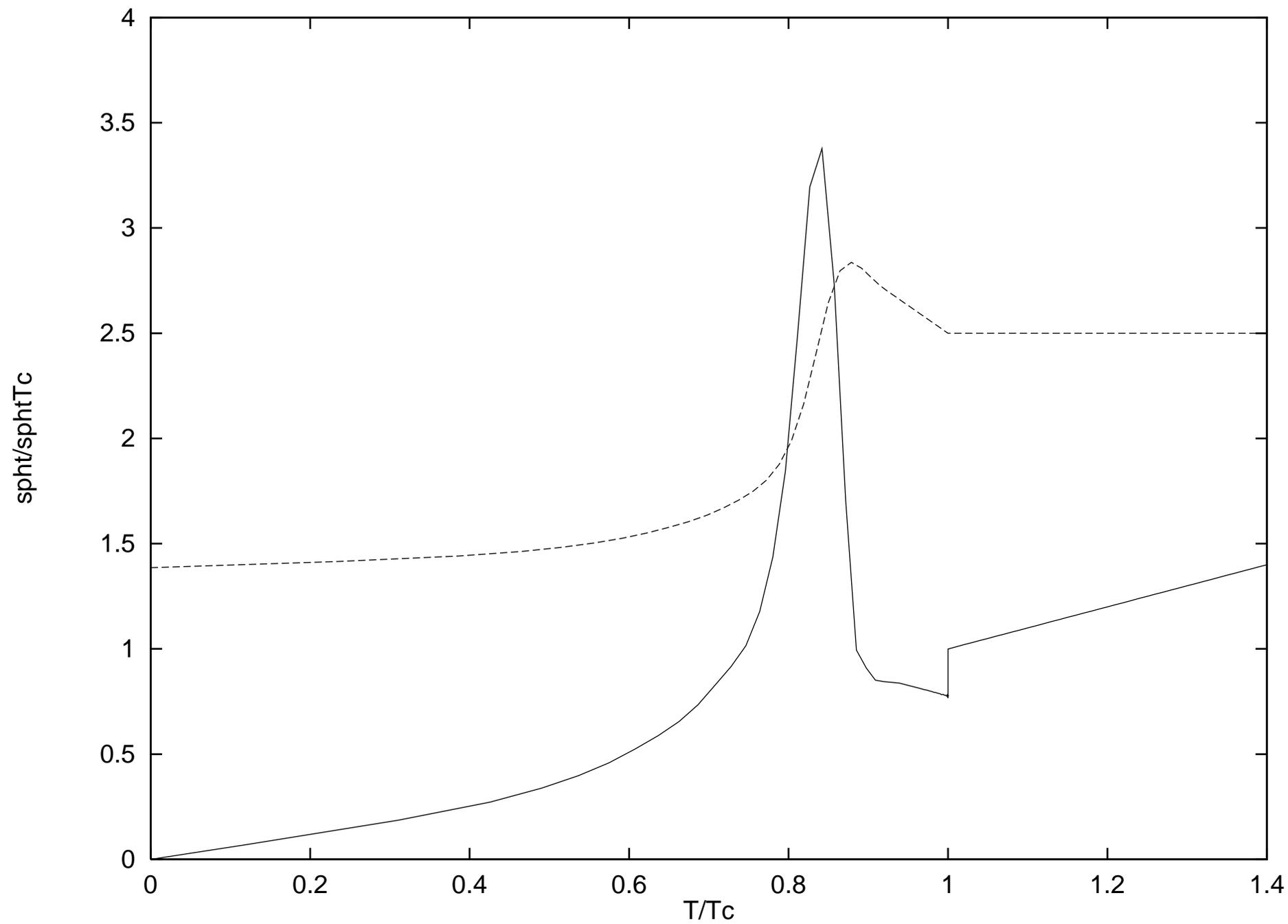
\end{document}